\documentclass[aps,prb,twocolumn,preprintnumbers,superscriptaddress,amsmath,amssymb]{revtex4-1}

\usepackage{graphicx}
\usepackage{epsfig}
\usepackage{dcolumn}
\usepackage{bm}
\usepackage{color}
\usepackage{amsmath}
\usepackage{amsfonts}
\usepackage{amssymb}
\usepackage{epstopdf}

\def\be{\begin{equation}}       \def\ee{\end{equation}}
\def\bea{\begin{eqnarray}}      \def\eea{\end{eqnarray}}
\def\ba{\begin{array} }
\def\ea{\end{array} }
\def\bnum{\begin{enumerate} }
\def\enum{\end{enumerate}}

\def\=>{\Rightarrow}
\def\>{\rightarrow}

\def\eye2{Fathbb{I}}

\renewcommand{\>}{\rangle}

\newcommand{\LDA}{\text{LDA}}
\newcommand{\vdW}{\text{vdW}}
\newcommand{\RPA}{\text{RPA}}
\newcommand{\Corr}{\text{Corr}}

\newcommand{\degree}{^{\circ}}

\begin{document}
\title{Moir\'e pattern interlayer potentials 
in van der Waals materials from random-phase approximation calculations}

\author{Nicolas Leconte}
\affiliation{Department of Physics, University of Seoul, Seoul 02504, Korea}

\author{Jeil Jung}
\email{jeiljung@uos.ac.kr}
\affiliation{Department of Physics, University of Seoul, Seoul 02504, Korea}

\author{S\'ebastien Leb\`egue}
\affiliation{
Laboratoire de Cristallographie, R\'esonance Magn\'etique et Mod\'elisations (CRM2, UMR CNRS 7036), Institut Jean Barriol, Universit\'e de Lorraine, BP 239, Boulevard des Aiguillettes, 54506 Vandoeuvre-l\`es-Nancy, France
}
\author{Tim Gould}
\affiliation{
Qld Micro- and Nanotechnology Centre, Griffith University, Nathan, Qld 4111, Australia
}

\begin{abstract}
  {
Stacking-dependent interlayer interactions are important for understanding the structural 
and electronic properties in incommensurable two dimensional material assemblies where 
long-range moir\'e patterns arise due to small lattice constant mismatch or twist angles.
Here, we study the stacking-dependent interlayer coupling energies
between graphene (G) and hexagonal boron nitride (BN) homo- and
hetero-structures using high-level random-phase approximation (RPA)
{\it ab initio} calculations. Our results show that although total
binding energies within LDA and RPA differ substantially between a factor of 200\%-400\%, 
the energy differences as a function
of stacking configuration yield nearly constant values with variations smaller than 20\% 
meaning that LDA estimates are quite reliable. 
We produce phenomenological fits to these energy
differences, which allows us to calculate various properties of
interest including interlayer spacing, sliding energetics, pressure
gradients and elastic coefficients to high accuracy.
The importance of long-range interactions (captured by RPA but not
LDA) on various properties is also discussed. Parameterisations for
all fits are provided.
}
\end{abstract}

\maketitle

\section{Introduction}


The quest for new artificial materials by assembling atomically thin two-dimensional
van der Waals materials~\cite{Koma_1992, Koma_1999, Geim:hf} has seen a new surge of interest during the last 
decade since the seminal transport experiments on graphene~\cite{Novoselov26072005, Novoselov:2005es, Zhang:2005gp}.
Artificial layered materials often form incommensurable crystals due to finite twist angles or differences in the lattice 
constants which leads to moir\'e patterns that dictate the appearance of a superlattice on top of the constituent crystal lattices.
These moir\'e patterns that form at the interface of incommensurable 
crystals lead to important features in the electronic structure of graphene 
at energy regions accessible by gate doping for sufficiently long moir\'e periods~\cite{Dean_2013,Ponomarenko_2013} 
opening up new possibilities of tailoring electronic properties through the control
of interface superlattices.
At the same time, non-negligible effects of moir\'e strains that reconfigure the stacking arrangement 
of the lattices in the 
limit of long moir\'e periods have been observed through tunnelling electron microscopy \cite{Alden_2013, Butz_2013},
and atomic force microscopy~\cite{Woods_2014}, rationalized by 
the quadratic decrease of the elastic energy with the moir\'e period~\cite{Jung_2015}.
Because the atomic and electronic structure of incommensurable moir\'e patterned  systems can be described as a collection of 
commensurate crystals with varying stacking configurations~\cite{Jung:2014ab},
an important first step towards understanding the physics of the moir\'e patterns 
is to understand the stacking dependent interlayer coupling between commensurate 
vertical heterolayer systems with short crystalline periods.
 
Two important examples of atomically thin van der Waals materials are graphene \cite{Geim:2007hy,RevModPhys.83.407,Geim:2007vw,CastroNeto:2009cl}, 
a single-atom thick sheet of carbon atoms arranged in a honeycomb lattice, 
and hexagonal boron nitride (BN) sheets~\cite{Dean_2010} whose 
honeycomb lattice consist of alternating boron and nitrogen atoms.
Graphene is a zero band gap semi-metal near charge neutrality that obeys a Dirac-like dispersion, whereas BN is a 
wide band gap insulator with an experimental bulk bandgap of 5.8 eV~\cite{PhysRevB.13.5560, PhysRevLett.53.2449}. 
Hexagonal boron nitride has been highlighted as an excellent dielectric barrier material
in field effect transistors with improved device mobilities through elimination of extrinsic factors like charged impurities and 
substrate ripples that limit the sample quality of graphene on conventional SiO$_2$ substrates~\cite{Dean_2010}.
This qualitative improvement in device qualities based on crystalline smooth barrier materials have led to the observation of 
new states of matter sensitive to disorder strength including new graphene fractional quantum Hall states~\cite{Du_2009,Bolotin_2009}, 
Fermi velocity renormalization~\cite{Elias_2011} and anomalously large magneto-drag~\cite{Gorbachev_2012}.
%
%
%
%
%
\begin{figure}
\includegraphics[width=\linewidth]{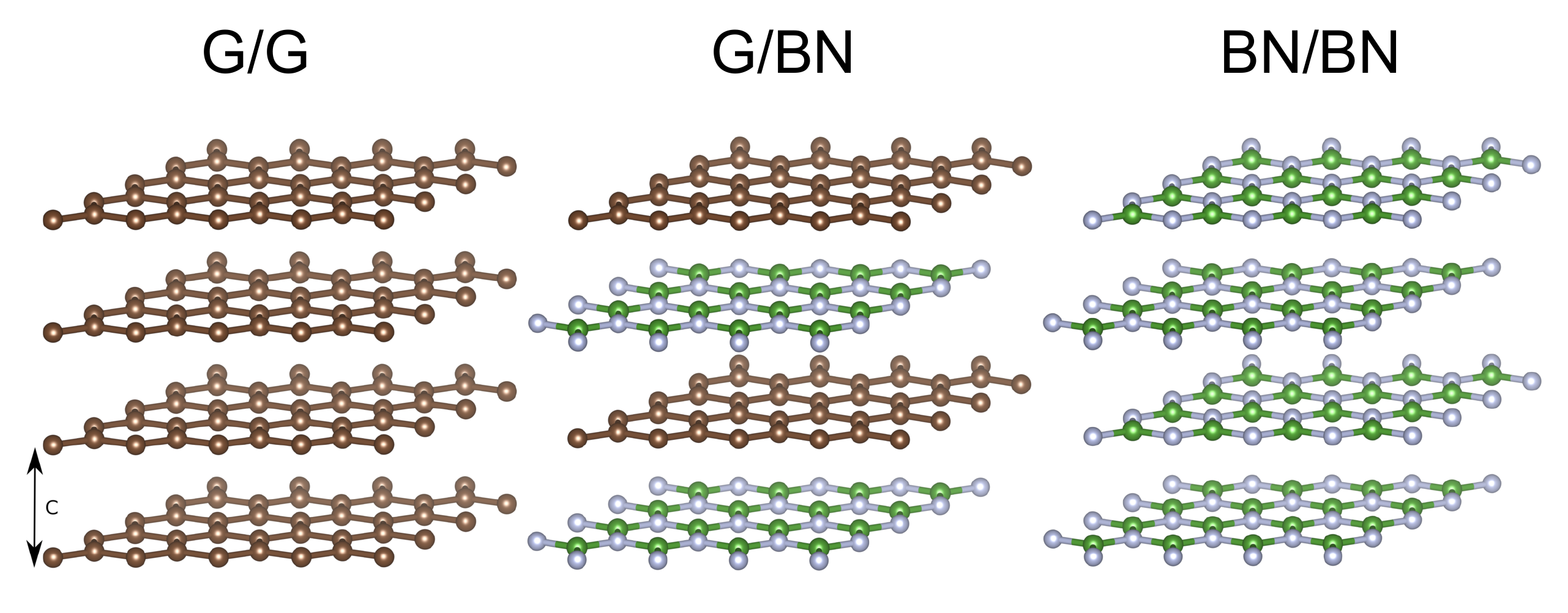}
\caption{
Sketch of the three bulk systems considered in this paper. Here, brown color corresponds to carbon forming graphene, while boron and nitrogen are gray and green, respectively. The interlayer spacing is represented by $c$. 
Two families of stacking configurations for 0$^{\circ}$ and $180^{\circ}$ in an aligned BN/BN system, 
as explained in the main text and illustrated in Fig.~\ref{fig:Configurations}.
}
\label{fig:system}
\end{figure}
By forming different elementary combinations of both materials (see Fig.~\ref{fig:system}), we can obtain graphene on graphene (G/G), 
mainly in its Bernal~\cite{Yan_2011, Kim_2016, Lin_2013} (AA-stacking is metastable~\cite{de_Andres_2008}) or twisted configuration~\cite{Brown_2012, Park_2015}, 
graphene on hexagonal boron nitride (G/BN)~\cite{Yankowitz_2012}, 
and boron nitride on boron nitride (BN/BN), 
that can form moir\'e superlattices whenever there is a lattice constant mismatch or finite twist angle. 
Recent experimental \cite{Yankowitz_2012,Hunt_2013,Woods_2014} and theoretical works \cite{PhysRevB.76.073103,PhysRevLett.99.256802,PhysRevB.86.155449,Bistritzer_2011,PhysRevB.87.245408,Jung_2015}
have noted the relevance of moir\'e patterns and moir\'e strains in configuring the electronic structure near charge neutrality and at energy scales close to the superlattice Brillouin zones corners.

In this work we calculate the interlayer interactions through a calculation of distance and stacking-dependent energy differences
that are required inputs to study the structural mechanics of the moir\'e strains in incommensurable crystals. 
This is a challenging task as the complex binding physics of layered van der Waals materials
 require theories that can explicitly account for the many-body effects~\cite{Dobson2012-JPCM,Gould2016-Chapter}. 
%
%
We present an accurate parametrization of the interlayer coupling energies between layered materials 
consisting of graphene and hexagonal boron nitride vertical heterostructures, including their 
dependence on interlayer stacking configuration difference. 

For high accuracy, total energies are calculated using high-level exact exchange and random phase approximation 
for the correlation energy (EXX+RPA or just RPA in short) \textit{ab initio} calculations
that are presented as a fitted correction to lower level local density approximation (LDA) calculations. 
The RPA is believed to be a good systematic approach to capture
the total energy differences for graphite\cite{PhysRevLett.105.196401} and other layered systems\cite{Bjorkman_2012}
We then use the fitted models to: i) Show that the LDA can serve as a solid backbone to estimate 
such energy differences and associated force-fields~\cite{Leven_2014} at reasonable computational cost. 
We note that for G/BN the Lennard-Jones types of pairwise potentials can grossly underestimate the 
stacking-dependent energy barriers~\cite{Neek-Amal:2014aa}
by almost an order of magnitude with respect to {\em ab initio} approaches \cite{PhysRevB.84.195414,Jung:2014ab}. 
Therefore, our calculations can provide a more reliable input for molecular dynamic codes to study, for instance, 
the friction between such layered 
materials~\cite{van_Wijk_2013, Reguzzoni_2012, Kitt_2013, Reguzzoni_2012a, Balakrishna_2014}. 
ii) Improve qualitative predictions for equivalent bilayer systems, for
sake of better experimental relevance. For this we use our fits to
approximate high-level RPA data for bilayer systems, for which
sufficiently accurate numerical RPA data is yet to be made available.

The rest of our manuscript is structured as follows. 
Section~\ref{methodology} focuses on the details of the methodology,
Sect.~\ref{results} discusses the results obtained for our different
systems while Sect.~\ref{summary} summarizes our findings.


\section{Methodology and computational details}
\label{methodology}

The methodology we use to obtain the interlayer interaction for the different possible 
G/G, G/BN, BN/BN heterojunctions draws from the {\em ab initio} theory of 
moir\'e superlattices~\cite{Jung:2014ab,Jung_2015} for incommensurable crystals where the local 
interlayer interaction is modelled based on calculations performed for short period commensurate geometries.
Similar earlier work attempting to capture interlayer interactions from different stacking geometries 
in commensurate G/BN were also presented in Refs.~[\onlinecite{PhysRevB.76.073103,PhysRevB.84.195414}].
From information at a few selected stacking configurations obtained from small unit cell 
commensurate calculations
we can build the energy landscape variations in the longer moir\'e pattern length scale
for different interlayer separation distances. 
Here we revisit the calculations for G/G~\cite{PhysRevLett.105.196401},
for BN/BN~\cite{PhysRevLett.96.136404,Bjorkman_2012} 
and G/BN heterostructures~\cite{PhysRevB.84.195414,Leven:2016aa,Thurmer:2017aa}, 
to analyze the stacking and interlayer distance dependent total energies 
in a consistent manner. 
All calculations are carried out with the \textit{ab initio} planewave
code VASP~\cite{PhysRevB.54.11169} for bulk systems. 
For RPA correlation energy calculations, 
we use an $8\times 8\times 3$ $\Gamma$-centered k-grid,
an energy cutoff of 700~eV, and a cutoff for the polarisability matrices of 300~eV.
For the Hartree-Fock energy calculations that provides the exact-exchange (EXX) energies, 
we use the same energy cutoff      
but increase the k-grid to $12\times 12\times 6$.
The LDA calculations use an energy cutoff of 500~eV and
a $\Gamma$-centered k-grid of $16\times 16\times 8$.
We use in-plane lattice parameters of
2.46~\AA~for graphene~\cite{PhysRevLett.105.196401},
2.50~\AA~for BN\cite{Pease_1952} and their average 2.48~\AA~
for the mixed G/BN system~\cite{PhysRevB.84.195414}.
With these parameter choices, our results for bulk hexagonal BN in the lowest energy AA' and AB configurations agree well with those
found in previous work~\cite{PhysRevLett.96.136404,Bjorkman_2012,PhysRevLett.111.036104}.
For example, for AA', we find an interlayer distance of $3.36$~\AA~versus $3.34$~\AA~from Ref.~[\onlinecite{PhysRevLett.111.036104}].
For the binding energy of AB, we get 42~meV/atom versus
39~meV/atom from Ref.~[\onlinecite{Bjorkman_2012}]. 
Results for G/BN are also similar to bilayer calculations reported in Ref.~[\onlinecite{Zhou2015}].

\begin{figure}
  \includegraphics[width=7cm]{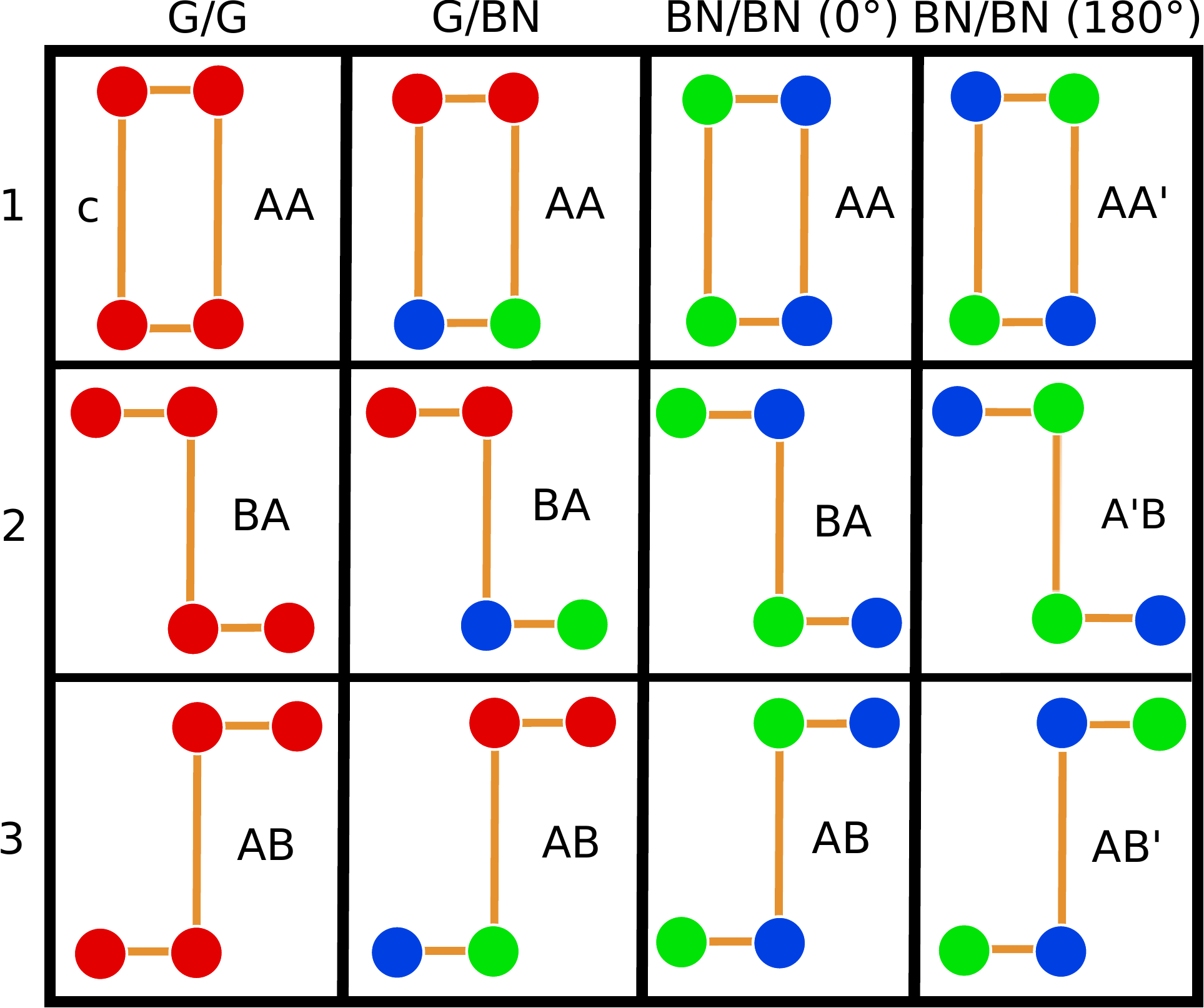}
  \caption{
  Definition of stacking configurations $S=1,2$ or $3$ for each system, in side view (top layer on top). Distance between layers is given by $c$. Carbon is red, boron is blue and nitrogen is green. For G/BN, we follow the definition in Ref.~[\onlinecite{Jung_2015}] where AB stacking denotes N on top of C ($a/ \sqrt{3}$), while BA corresponds to B on top of C ($2a /\sqrt{3}$). 
  For the other BN/BN configurations, we match naming conventions with the ones in Ref.~[\onlinecite{PhysRevLett.111.036104}].
  By using the energies associated with each of these positions (or any other combination of three positions), one can extract the potential landscape of all stacking configurations.}
    \label{fig:Configurations}
\end{figure}

To accurately interpolate the RPA
results~\cite{PhysRevB.13.4274, PhysRevB.15.2884, Langreth19751425} as
a function of interlayer separation distance $c$,
we use the scheme suggested in Ref.~[\onlinecite{Gould_2013}].
We approximate RPA results by correcting LDA energies using
\begin{align}
  U^{\RPA}_{S}(c)\approx U^{\LDA}_{S}(c) + U_{\Corr}(c).
  \label{eqn:UFit}
\end{align}
Here $S$ denotes the chosen stacking configuration, see
Fig.~\ref{fig:Configurations} for an illustration of the corresponding configurations.

This approach takes advantage of the good \emph{short-range} accuracy
of LDA DFT, but corrects its poor treatment of \emph{long-range} effects
using RPA results.
By assuming that LDA is valid for distances below equilibrium
separation where short-range covalent-binding dominates, and that the
longer-range vdW dispersion potential takes the upper hand for
distances beyond the equilibrium distance, we can separate both
contributions estimating the correction term by
\begin{align}
U_{\Corr}=&f(c)[U_{\vdW}(c) - U^{\LDA}_{S}(c)]
\end{align}
where
\begin{eqnarray}
f(c)=[1 + \kappa_S \exp(-(a_1^S x_{S}+a_2^S x_{S}^2+a_3^S x_{S}^3)]^{-1},
 \label{eqn:f}
 \end{eqnarray}
and use for the van der Waals tail description the function
 \begin{eqnarray}
  U_{\vdW}(c)&=&-\frac{C_4}{(c^4-D_s^4)} - \frac{C_3}{c^3} \frac{2}{\pi} \arctan\left(\frac{c}{D_C} + \phi_c\right)
  \label{eqn:vdW}
  \end{eqnarray}
for graphite to account for the interaction between the Dirac cones in G/G and for consistency with the asymptotic behavior in Ref.~[\onlinecite{Gould_2013}].
For all other systems when we have an insulating gap we use    
 \begin{eqnarray}
U_{\vdW}(c)&=&-\frac{C_4}{(c^2-D_s^2)^2}.  
  \label{eqn:vdW2}
  \end{eqnarray}  
The LDA part is given by
  \begin{eqnarray}
  U^{\LDA}_{S}(c)=&-M_0^{S}\left[\frac{\tau^{S}_2 e^{-\tau^{S}_1 x_{S}^\text{LDA}} - \tau^{S}_1 e^{-\tau^{S}_2 x_{S}^\text{LDA}}}{\tau^{S}_2 - \tau^{S}_1}\right].
  \label{eqn:ULDA}
\end{eqnarray}
where $x_{S}=c/c_{RPA}^{S}-1$ and $x_{S}^\text{LDA}=c/c_\text{LDA}^{S}-1$. 
Eq.~(\ref{eqn:ULDA}) provides a fitting model for the LDA calculation of stacking S and simplifies to
\begin{align}
U_{\LDA}^{S}(c)=&-M_0^{S}(1 + \tau^{S} x_{S}^\text{LDA})e^{-\tau^{S} x_{S}^\text{LDA}}
  \label{eqn:ULDA2}
\end{align}
when $\tau^{S}_1 = \tau^{S}_2 = \tau^{S}$.
This fitting approach allows us to closely compare the RPA results with LDA (or any other approximation) values as a function of 
different interlayer separation. 

Furthermore, this fitting offers a second advantage. Due to the high
computational cost for carrying out calculations for bilayer systems where a large vacuum is required, we can presently only obtain
reliable RPA data for bulk systems. 
Using this fitting procedure it is possible to extract the parameters that approximate the behavior of bilayer systems
using LDA calculations for bilayers and fitting again the parameters using the 
long-range correction terms estimated from the bulk behavior~\cite{Gould_2013}, see Appendix~\ref{app:bi} for a more detailed discussion. 
This procedure is used to obtain the modified bilayer fitting
parameters presented in Table~\ref{dataTable} 
to obtain estimates for the total energy curves in bilayer geometries at RPA-level accuracy.
%
%
%
%
%
%

By calculating the bulk quantities for three stacking configurations, a general behavior of the interlayer 
binding energies can then be extrapolated for every case based on the approach 
outlined in Ref.~[\onlinecite{Jung_2015}].
The stacking-dependent energy landscape, in the first harmonic approximation, 
is given by
\begin{equation}
  U(x,y,c) \approx C_0(c) + f_1(x,y,c,C_1,\phi_0)
  \label{landscape} 
\end{equation}
where $x$, $y$ are the in-plane stacking coordinates and $c$ is the interlayer separation.
The function $f_1$ follows from trigonal symmetry and is defined as
\begin{multline}
  f_1(x,y,c,C_1,\phi_0) = 2 C_1 \cos(\phi_0 - G_1 y)
  \\ + 4 C_1 \cos(G_1 y/2 + \phi_0) \cos(\sqrt{3}G_1 x/2).
  \label{eq:f1}
\end{multline}
where $C_0$, $C_1$ and $\phi_0$ are the three parameters to be fitted and $G_1 = 4 \pi /\sqrt{3} a$ is the magnitude of the reciprocal lattice vector.
 In the case we have information of AA, AB and BA stacking configurations these $c$ dependent parameters can be written as follows~\cite{Jung_2015}
\begin{equation}
  \phi(c) = \arctan\left[ - \frac{\sqrt{3}}{2(D+1/2)}\right],
  \label{eq:phi}
\end{equation}
\begin{equation}
  C_1(c) = \frac{U(0,2/\sqrt{3},c)-U(0,1/\sqrt{3},c)}{6 \sqrt{3} \sin(\phi(c))}
  \label{eq:C1}
\end{equation}
and
\begin{equation}
  C_0(c) = -6 C_1 \cos(\phi(c)) + U(0,0,c)
  \label{eq:C0}
  \end{equation}
where
\begin{equation}
D = \frac{U(0,0,c) - U(0,1/\sqrt{3},c)}{U(0,1/\sqrt{3},c)-U(0,2/\sqrt{3},c)}.
\end{equation} 
We also derive more general expressions in Appendix~\ref{app:param} that allow to combine \textit{any} three stacking configurations to parametrize the in-plane potential landscape. 

Finally, we calculate the interlayer elastic coefficient $C_{33}$ and the interlayer inelastic coefficient $C_{333}$ for the three stacking configurations of each system
as defined in Ref.~[\onlinecite{Gould_2013}] 
\begin{equation}
\frac{F_3(c)}{V_0} \approx C_{33} \left( \frac{c}{c_0} - 1 \right) +\frac{1}{2} C_{333} \left( \frac{c}{c_0} - 1 \right)^2 
\end{equation}
where the normalized force per unit volume
$F_3/V_0 \equiv (c / c_0) (dE/dc) $
depends on distortions in the out-of-plane direction through
\begin{equation}
  C_{33} = \frac{c_0^2}{V_0} \frac{d^2 E(c)}{dc^2} \Bigr|_{c_0}
  \label{C33}
\end{equation} 
and
\begin{equation}
  C_{333} = \frac{c_0^3}{V_0} \frac{d^3 E(c)}{dc^3} \Bigr|_{c_0}.
  \label{C333}
\end{equation}


\begin{figure}[t]
  \includegraphics[width=\linewidth]{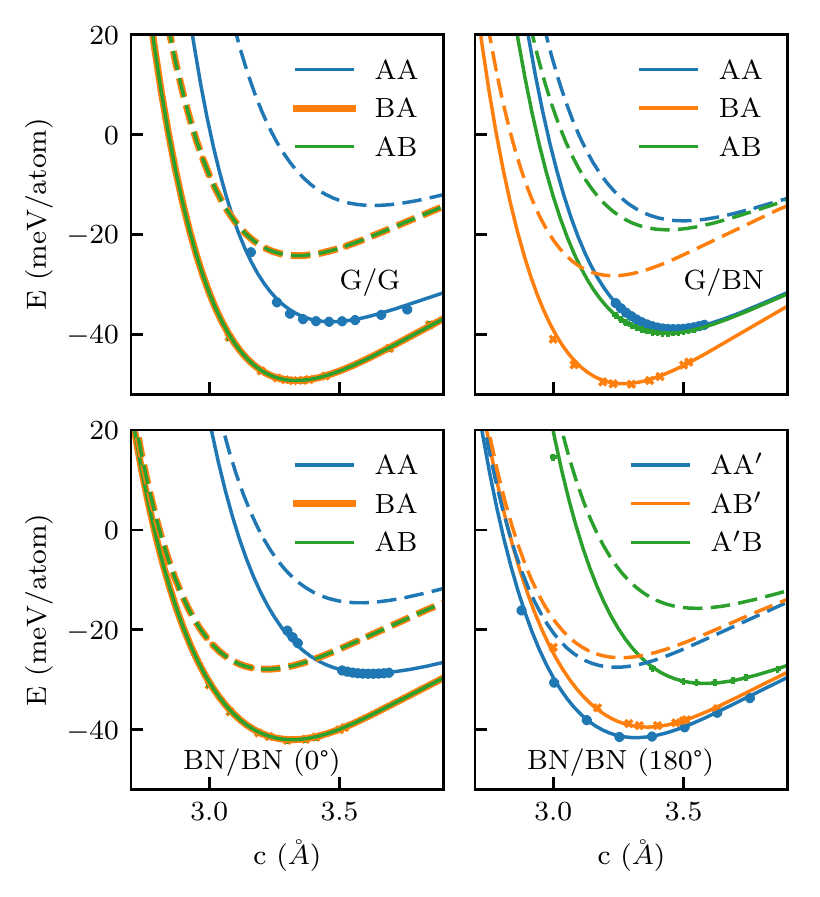}
  \caption{Total energy from accurate bulk RPA and LDA calculations, as well as corresponding fitting lines based on Eqs.~(\ref{eqn:UFit}) to (\ref{eqn:ULDA2}).
  for different stacking configurations for each system corresponding to parametrization given in Table~\ref{dataTable}, as defined in Fig.~\ref{fig:Configurations}. 
  The symbols are calculated data points while the fits are represented as lines.  
  The dashed lines are bulk LDA total energy fits. Our fitting procedure is particularly accurate in the region of interest where $c = 3\sim4 \AA$. }
    \label{fig:etot}
\end{figure}

%
\section{Results and discussions}
\label{results}

In this section we discuss the interlayer interaction energies obtained from the EXX+RPA calculations 
for the different G/G, G/BN and BN/BN heterostructures considered. 
The fitting scheme for the interlayer energy curves based on the Eqs.~(\ref{eqn:UFit}) to (\ref{eqn:ULDA2}) are 
illustrated in Fig.~\ref{fig:etot} where we show the fitted curves in solid lines together with the dataset represented 
by symbols for the different stacking configurations illustrated in Fig.~\ref{fig:Configurations}. 
When we approximate the bilayer RPA behavior (see Table~\ref{dataTable})
we obtain energies that are about twice as small as the bulk values (not shown here)
consistent with the fact that there are fewer interfaces.
The total energy values reported in this manuscript should be considered accurate to at best $1$~meV/atom due to 
uncertainties related with methodological errors in the extrapolation, and numerical convergence.

\begin{figure}[t]
  \includegraphics[width=\linewidth]{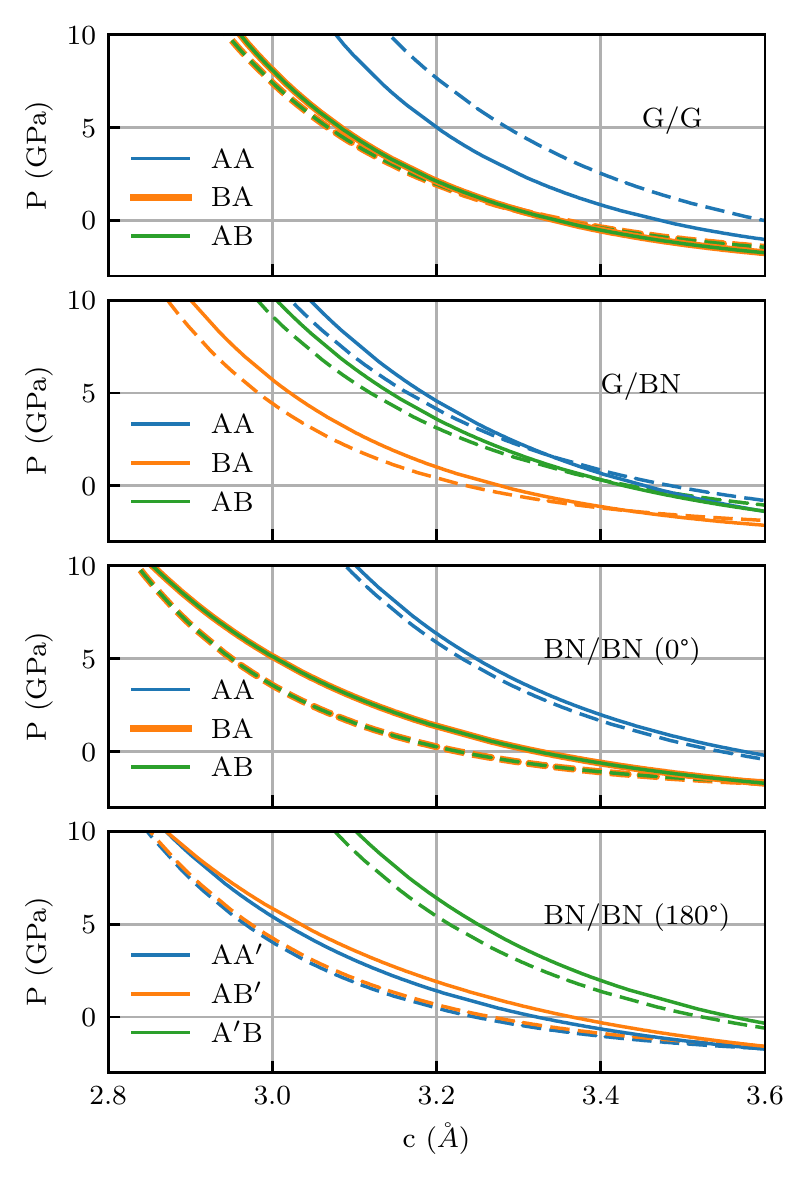}
  \caption{(Color online) Bulk equilibrium interlayer distance corresponding to different stacking positions as a function of pressure obtained at the RPA (solid) and LDA (dashed) level. 
 The different colors represent the different stacking configurations that are defined in Fig.~\ref{fig:etot}. 
 The gradients of both approximations are very similar in the compression regime and have maximum deviations for the predicted equilibrium distances of $\sim$0.1~$\AA$ in the worst cases.  
 Therefore, the LDA can be used as a reliable approximation for estimating the changes in interlayer distance with pressure.}
    \label{pressureFig}
\end{figure}

The pressure curves as a function of distance obtained by fitting the distance dependent energies
with a Birch-Murnaghan equation of state~\cite{PhysRev.71.809} are shown in Fig.~\ref{pressureFig} for different stacking configurations.
The results are provided both at the LDA (dashed lines) and RPA (solid lines) which show qualitative agreements in the ordering of the forces
for the different stacking configurations although there are quantitative differences. 

\begin{figure*}
  \includegraphics[width=16cm]{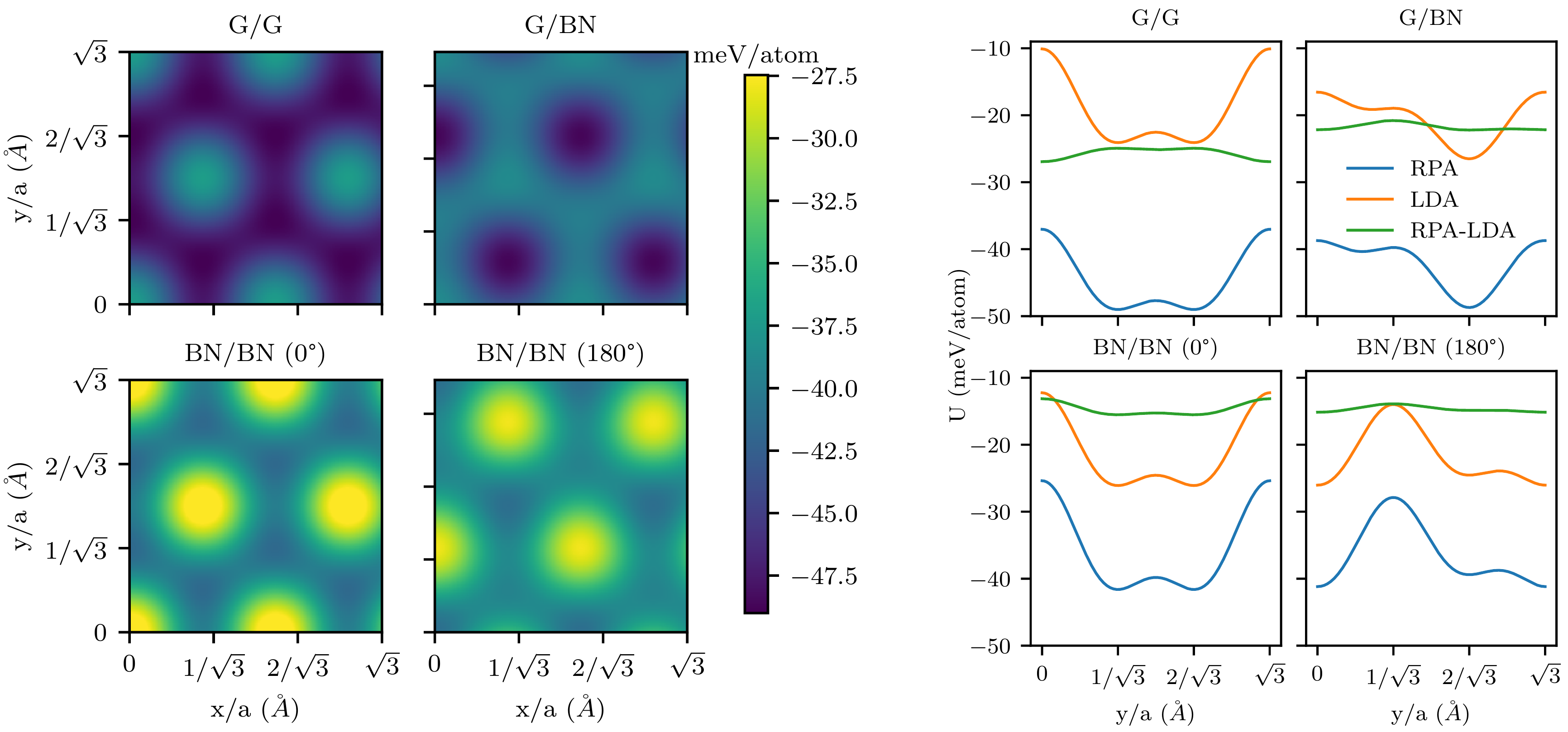}
  \caption{(Color online) {\em Left panel:} Two-dimensional maps of energy landscapes given by Eq.~(\ref{landscape}) for the RPA fits, for $c=3.4 \AA$. The parameters building these fits are in turn represented in Fig.~\ref{fig3}, as well as a cut of the energy map along the $x=0$ axis. The energy differences are largest for the BN/BN ($60\degree$) system of non-alternating atoms between layers, suggesting larger lattice reconstruction than for the other systems.
 { \em Right panel:}
  The vertical cut at $x=0$ of the energy landscape (at an interlayer distance of $3.4 \AA$) in Fig.~\ref{fig:maps}, for both LDA (orange) and RPA (blue) approximations, as well as their respective difference (green curve). The nearly constant behaviour of the latter supports the main message of the paper, namely that the LDA yields accurate predictions for any type of quantity that takes energy differences as input variable. For the G/BN and BN/BN ($0\degree$) system, the $y$-coordinates of AB (BA) stacking correspond to $a/{\sqrt{3}}$ ($2a/{\sqrt{3}}$), respectively where $a$ is the lattice constant of the unit cell. For the BN/BN ($180\degree$) system, the $y$-coordinates of A$^\prime$B (AB$^\prime$) stacking correspond to $a/{\sqrt{3}}$ ($2a/{ \sqrt{3}}$) respectively.
  }
    \label{fig:maps}
\end{figure*}

\begin{figure}
\centering
  \includegraphics[width=8cm]{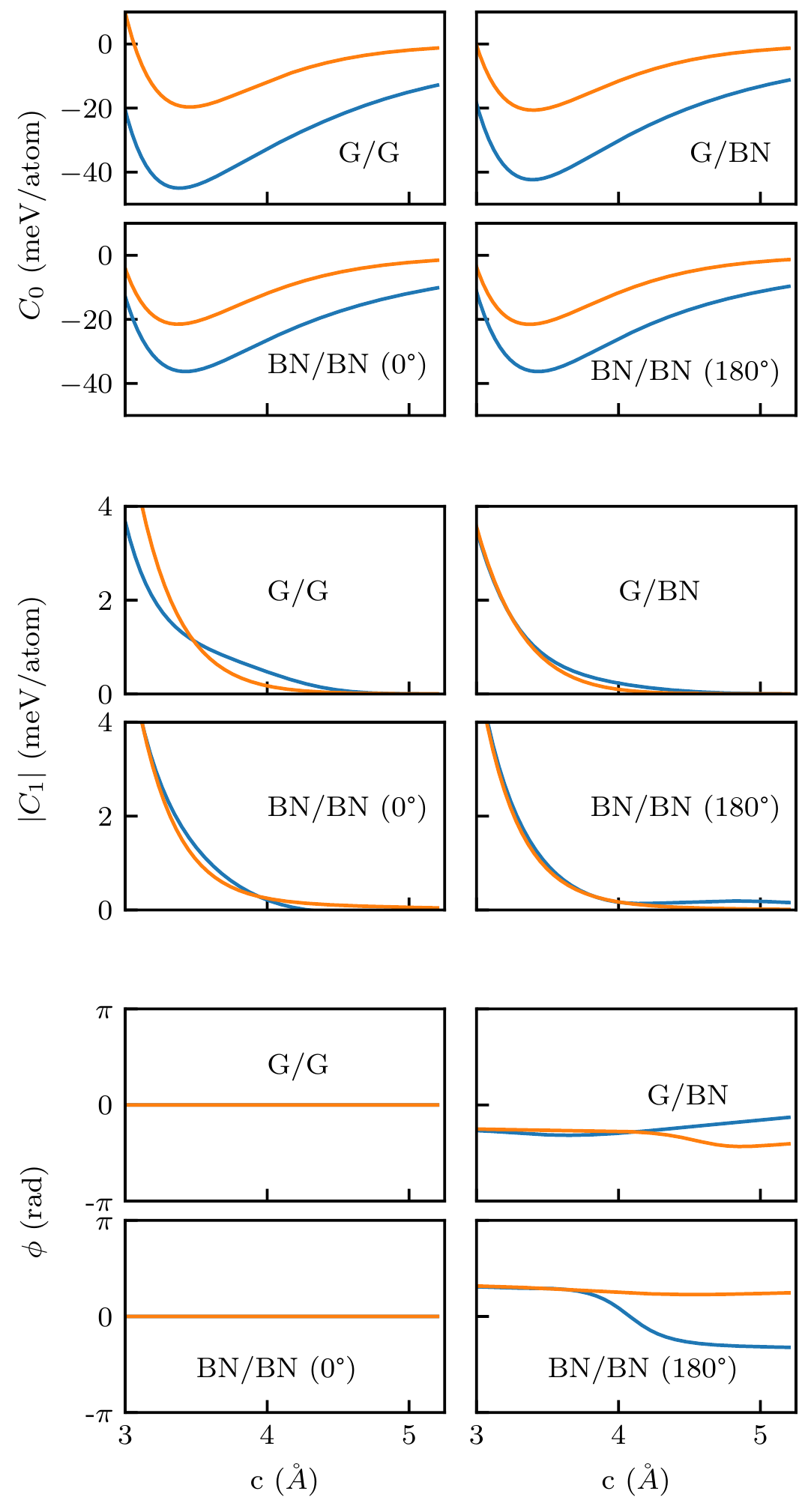}%
\caption{(Color online) 
The parameters in Eqs.~(\ref{eq:phi})-(\ref{eq:C0})
the  $C_0(c)$ captures the average stacking dependent total energy at a given interlayer separation $c$,
$C_1(c)$ the first Fourier component magnitude and $\phi(c)$  the phase  associated to the stacking dependent energy modulation landscape
given in the LDA (orange) and RPA (blue) approximations for each one of the systems considered.}
\label{fig3}
\end{figure}

The energy landscapes based on Eq.~(\ref{landscape}) representing the total energies for different stacking at a fixed interlayer distance of $c=3.4~\AA$ 
are shown in Fig.~\ref{fig:maps}. 
{
Using a shared colormap between the different systems it is possible to distinguish the contrasts in the total energies, we see that, as expected, the less stable BN/BN ($0\degree$) system produces the largest energy variations between different stackings (up to $\sim 16$ meV/atom), opposing smooth sliding between the layers and potentially enhancing in-plane moir\'e strains.
The other systems have comparatively smaller maximum energy differences: G/BN is lowest with $\sim 10$ meV/atom while BN/BN ($180 \degree$) and G/G systems generate values of about $13$ and $12$ meV/atom, respectively. In Fig.~\ref{fig3}, we plot the parameters $C_0(c)$, $C_1(c)$ and $\phi(c)$ that control this stacking dependent energy-landscapes, as given by Eqs.~(\ref{eq:phi}) to (\ref{eq:C0}) for each system. 
The  $C_0(c)$ is the average stacking dependent total energy at a given interlayer separation $c$,
whereas $C_1(c)$ and $\phi(c)$ are the magnitude and phase of the stacking dependent energy modulation described within the first harmonics.
The magnitude represents the amplitude of the oscillation while the phase indicates the degree of mixing between inversion symmetric and inversion asymmetric contributions to the moire pattern modulations.~\cite{hbnmoire}
The lower-right $2\times2$ panel gives the vertical cut at $x=0$ of the energy landscape for both LDA and RPA approximations and their differences. An overview of all the numerical data based on this procedure outlined in Sect.~\ref{methodology} is provided in Table~\ref{dataTable}.
}
Finally, the interlayer elastic and inelastic coefficients, given by Eqs.~(\ref{C33}) and (\ref{C333}), 
calculated at the equilibrium separation $c_0$ are summarized in Table~\ref{dataTable2}.

In the following we discuss in some detail the interlayer interaction properties of the different systems
consisting of G/G, G/BN and the two different BN/BN stacking configurations.

\subsection{G/G}
%
%

The interlayer binding energy of bilayer graphene can be understood as the elementary cohesive energy between the layers in graphite. 
The cleavage energy, approximately equal to the binding energy, of graphite has been measured based on the self-retraction phenomenon in graphite~\cite{Wang_2015, PhysRevLett.100.067205},
while computationally the cohesive energies have been calculated in the past at different levels of approximation~\cite{Grimme_2006, PhysRevB.82.081101, Sun_2013, PhysRevB.21.5469, Sorella_2007, PhysRevLett.99.166401},
and more recently through accurate RPA calculations carried out on graphite~\cite{PhysRevLett.105.196401} 
that allowed to confirm the weak non-additivity effects due to long-range van der Waals interactions. 
Within RPA the binding energies at the equilibrium distance are equal to $49$ meV/atom at Bernal stacking and $36$ meV/atom for the least stable AA stacking (see Fig.~\ref{fig:etot}),
while for intermediate stacking configurations the binding energies vary between these two values as shown in Fig.~\ref{fig:maps}.   

The elastic and inelastic coefficients listed in Table~\ref{dataTable2} (also for AA stacking, extending the available data for Bernal stacking~\cite{Gould_2013}) are 
significantly enhanced (up to $40 \%$) when the long-range interactions are included within RPA compared to the LDA. 
For the bilayer coefficients one obtains values that are of the same order of magnitude as the bulk when we multiply the results by two
(we do not report the inelastic coefficients of bilayer RPA, as the results are only approximate and we cannot benchmark it against directly calculated RPA data yet). 
This factor two multiplication is required to make a comparison with the bulk as there are twice as many interlayer neighbors in the latter case.

The energy profile for G/G resulting from the fitting parameters are plotted in Fig.~\ref{fig3}. The $C_0$ corresponds to the average between the energies at the $AA$, $AB$ and $BA$ stacking, 
while the binding energy equal to $44$ meV/atom is a value that is more than doubled when compared to the LDA. 
The differences between the energy average and the minimum is approximately 4~meV/atom and indicates the order of magnitude for the energy gradient that controls the in-plane forces~\cite{Jung_2015}.    
The relatively flat green curve (based on the difference between RPA and the LDA absolute energy data) in the lower-right panel of the figure illustrates
that LDA yields accurate predictions on energy differences for this system that are fairly close to the RPA results.

\subsection{G/BN}


When we calculate the total energies for graphene and BN heterojunctions, we ignore the 
$\sim$2~$\%$ lattice constant mismatch and obtain the interlayer stacking-dependent total energies
as in Ref.~[\onlinecite{PhysRevB.84.195414}] using an averaged lattice constant of $a=2.48~\AA$. 
These stacking dependent total energies based on LDA calculations were useful references for 
identifying the role of spontaneous strains in G/BN heterojunctions giving rise to a band gap~\cite{Jung_2015}.     
The fitted RPA results for different stacking and interlayer distances are plotted in Fig.~\ref{fig3} where the green curve in the lower 
right panel validates the use of LDA data to estimate the stacking-dependent energy differences and associated strains in Ref.~[\onlinecite{Jung_2015}].
The total energy difference between the least favorable AA and most favorable BA stacking configuration is of the order of 10~meV/atom and is comparable to the LDA results,
as well as the stacking dependent total energy differences in G/G.
%
%
%
Our binding energy of $23$ meV/atom estimated from bulk is in fair agreement with the direct calculation of 
$21$ meV/atom in the isolated bilayer geometry in Ref.~[\onlinecite{PhysRevB.84.195414}].

When we calculate and compare the interlayer elastic and inelastic coefficients, we observe, similarly to the G/G system, 
a drastic enhancement when including long-range corrections as compared to the LDA calculations, up to $40 \%$ for bilayer AA stacking
and therefore the use of the RPA data is required to properly estimate these constants. 
The largest elastic coefficients are obtained at the most stable BA structure that corresponds to the situation where one carbon atom is on top of boron.


\subsection{BN/BN}

%

\begin{table*}[] 
\begin{footnotesize}
    \begin{tabular}{ | c c | c c c | c c c | c c c | c c c|}
      \hline
         &  &  & G/G &  &  & G/BN &  &  & BN/BN ($180\degree$) & &  & BN/BN ($0\degree$) & \\ 
         \hline
         & Configuration (S)  & AA & BA & AB & AA & BA & AB & AA$^\prime$ & AB$^\prime$  &  A$^\prime$B & AA & BA & AB \\ 
         \hline
        $\text{LDA}_{\text{Bulk}}$ & $M_0^S$ & $14.23$ & $24.271$ & $24.271$ & $17.3$ & $28.3$ & $19.1$   & $27.5$  & $25.6$  & $15.7$  & $14.6$  & $27.9$  & $27.9$    \\ 
         & $c^\text{LDA}_S$ & $3.631$ & $3.341$  & $3.341$ & $3.5$  & $3.23$ & $3.44$   & $3.24$  & $3.26$  & $3.55$  & $3.58$ & $3.22$  & $3.22$    \\ 
         & $\tau_1^S$ & $9.373$ & $8.412$ & $8.412$ & $8.699$ & $6.541$ & $8.52$   & $7.855$ & $7.886$ & $7.828$ & $7.736$ & $6.0$  & $6.0$   \\  
         & $\tau_2^S$ & $9.373$ & $8.412$ & $8.412$ & $8.699$ & $10.177$ & $8.52$  & $7.855$ & $7.886$ & $10.621$ & $10.836$ & $10.0$  & $10.0$  \\
                  \hline
        $\text{LDA}_{\text{Bi}}$ & $M_0^{S}$ & $9.664$ & $13.312$ & $13.312$ & $8.232$ & $14.110$ & $9.155$   & $13.832$  & $12.500$  & $7.622$  & $7.090$  & $13.809$  & $13.809$    \\ 
          & $c^\text{LDA}_{S}$ & $3.557$ & $3.32$  & $3.32$ & $3.535$  & $3.216$ & $3.457$   & $3.24$  & $3.3$  & $3.56$  & $3.6$ & $3.25$  & $3.25$    \\ 
         & $\tau_1^{S}$ & $8.645$ & $7.837$ & $7.838$ & $8.181$ & $7.569$ & $8.058$   & $7.827$ & $8.399$ & $9.356$ & $8.610$ & $8.123$  & $8.123$   \\  
         & $\tau_2^{S}$ & $8.645$ & $7.837$ & $7.838$ & $8.181$ & $7.569$ & $8.059$  & $7.827$ & $8.399$ & $9.356$ & $8.610$ & $8.124$  & $8.124$  \\

         \hline
        $f$ & $\kappa_S$ & $1.262$ & $1.373$ & $1.373$ & $1.038$ & $1.324$ & $1.17$ & $1.756$ & $1.674$ & $1.188$  & $1.254$ & $1.699$ & $1.699$  \\ 
         & $a_1^S$ & $6.843$ & $11.496$ & $11.496$ & $10.8$ & $12.7$ & $11.1$      & $13.5$  & $14.0$ &  $12.3$   & $12.0$  & $14.2$& $14.2$    \\ 
         & $a_2^S$ & $3.315$ & $-5.586$ & $-5.586$  & $-11.8$ & $-11.8$ & $-9.8$   & $-24.5$ & $-35.0$ & $-19.7$   & $-3.8$ & $-28.3$ & $-28.3$  \\  
         & $a_3^S$ & $30.0$ & $30.0$ & $30.0$ & $30.0$ & $30.0$ & $30.0$ &          $30.0$    & $30.0$ &  $30.0$   & $30.0$ & $30.0$ & $30.0$   \\ 
         & $c^\text{RPA}_S$ & $3.476$ & $3.334$ & $3.334$ & $3.46$ & $3.27$ & $3.43$ &          $3.32$    & $3.36$ &  $3.58$   & $3.62$ & $3.32$ & $3.32$   \\ 
         \hline
         vdW & $C_4$ &  & $7570$ & &  & $7800$ &  & & $7100$ &  & & $7100$ &   \\ 
         & $C_4^\text{Bi}$ &  & $3492.72$ & &  & $3603.6$ &  & & $3280.2$ &  & & $3280.2$ &   \\
         & $D_S$ & & $2.22$ &   &  & $0.86$ &   & & $0.86$ &   &  & $0.86$  & \\ 
         & $C_3$ & & $380$ &  &  & $0$ &     &   & $0$ &    &  & $0$&     \\ 
         & $C_3^\text{Bi}$ & & $172.9$ &  &  & $0$ &     &   & $0$ &    &  & $0$&     \\ 
         & $D_C$ &  & $23.7$ & & & $0$ & &      & $0$ &   &  & $0$ &    \\ 
         & $\phi_c$ &  & $0.62$ & & & $0$ &  &     & $0$ &  &  & $0$ &    \\ 
         \hline
          \end{tabular} 
          \end{footnotesize}
    \caption{
      Summary of numerical data based on the procedure outlined
      in Sect.~\ref{methodology}, as given by Eqs.~(\ref{eqn:f}),
      \eqref{eqn:vdW}/\eqref{eqn:vdW2} and \eqref{eqn:ULDA}.
      We differentiate between
      parameters that reproduce the LDA calculation, the vdW
      correction and the fitting function $f$. The $C_3$ term exists
      only for systems with interactions between Dirac modes in
      different layers. 
      For the bilayer systems we need to use a new set of LDA parameters to obtain the fits as well as modified 
      $C_3^\text{Bi}$,  $C_4^\text{Bi}$ parameters  (see Appendix~\ref{app:bi} for details). 
      We note that the interlayer distances $c^\text{LDA}_S$ are different between the bulk 
      and the bilayer systems, which we rationalize by the fact that a single layer in bulk
      is surrounded on both sides of the layer while for the bilayer
      the interface is only on one side.} 
     \label{dataTable}
\end{table*}

Hexagonal boron nitride layers share many similar aspects to the bilayer graphene while
the most notable difference is the polar character of their interatomic bonds and the marked
distinction between each atom species within each layer. 
Due to their ionic character, the most
stable crystalline form in their hexagonal geometry is the vertically alternating arrangement of the
atoms in the ${\rm AA}'$-stacking configuration (in our figures and table referred to as BN/BN $180\degree$).
We also provide data for the case with non-alternating atoms (BN/BN $0\degree$). 
We note that according to our RPA data the AB configuration is nearly as stable as the AA$^\prime$ one 
(less than $1$ meV/atom), thus explaining the existence of both configurations in experiment~\cite{Warner_2010}.

The resulting fitting parameters for these BN/BN systems are plotted in Fig.~\ref{fig3} and confirm our main 
conclusions regarding the qualitative validity of LDA data. 
We further note that the BN/BN systems give larger values of $C_1$, indicating that these system will have a stronger tendency to lock into an
energetically more stable stacking configuration.

Unlike the G/G and G/BN systems, in BN/BN systems the LDA and RPA predict similar interlayer elastic and inelastic coefficients, perhaps reflecting a greater role for ionic effects that are well-captured by LDA.
Nevertheless, small changes are still observed and one should resort to RPA data whenever available. 
%


\begin{table*}[] 
    \begin{tabular}{ | c c | c c c | c c c | c c c | c c c|}
      \hline
         &  &  & G/G &  &  & G/BN &  &  & BN/BN ($180\degree$) & &  & BN/BN ($0\degree$) & \\ 
         \hline
         & Configuration (S)  & AA & BA & AB & AA & BA & AB & AA$^\prime$ & AB$^\prime$ &   A$^\prime$B& AA & BA & AB \\
         \hline
	& $C_{33}^\text{RPA}$ & 29  & 37  & 37  & 32  & 38  & 30  & 32  & 32 & 24 &18 & 31 & 31 \\ 
         & $C_{333}^\text{RPA}$ & -580 & -600  & -600  & -560 & -570 & -500  & -420 & -370  & -430 & -360 & -370 & -370 \\ 
         & $C_{33}^\text{LDA}$ & 21 & 31 & 31  & 23  & 35& 24 & 31 & 29 & 21 & 20 & 31 & 31 \\ 
         & $C_{333}^\text{LDA}$ & -400  & -520 & -520  & -400 & -580 & -420  & -480 & -450  & -400 & -370 & -490 & -490 \\ 
         & $C_{33}^\text{RPA,Bi}$ & 17 & 18  & 18  & 15 & 18 & 14 & 16 & 18  & 16 & 10 & 17 & 17 \\ 
         & $C_{33}^\text{LDA,Bi}$ & 12  &  15 & 15 & 9 & 15  & 10  & 16 & 16  & 11 & 9 & 17 & 17 \\ 
         \hline
    \end{tabular} 
    \caption{
    Summary of the interlayer elastic $C_{33}$ and inelastic $C_{333}$ coefficients as 
    defined in Eqs.~(\ref{C33}) and (\ref{C333}), in [GPa], comparing values obtained within LDA and RPA, for bulk and bilayer systems. 
    The $C_{333}$ values are not listed for bilayers because we have not carried out direct bilayer RPA calculations.
    The stacking configurations $1$, $2$ and $3$
      are represented in Fig.~\ref{fig:Configurations}. 
      The relevance of long-range correlations is partly manifested in the impact made on 
      these constants by the RPA.
      }
     \label{dataTable2}
\end{table*}

\section{Summary and discussions}
\label{summary}

We have presented an accurate parametrization of the van der Waals
interaction energies in 2D artificial materials that can be formed
using graphene and hexagonal boron nitride single layers. 
Our methodology based on the RPA density-density response function is able
to capture from first principles the many-body non-local Coulomb
correlation effects that are responsible for a large part of
interlayer binding.

The benchmark against our first principles EXX+RPA
calculation suggests that the success of the LDA in calculating the
equilibrium geometries in the systems we considered  for different stacking can be traced to
its ability for capturing reliably the electronic structure in the
covalent regime where the interatomic repulsion is important, and to a
fortuitous tendency to overbind the layers at a moderately large
interlayer separation distance. 
The LDA can thus be considered an accurate first approximation to predict friction energies in the layered materials considered.

We note that advances for methods beyond the LDA have already been
made~\cite{Leven:2016aa} using implementations by Tkatchenko-Scheffler
and Many-Body Dynamics methods\cite{TS,MBD}
to account for the dispersion forces. 
However, despite ongoing improvements\cite{Gould_2016}
in the semi-empirical treatment of dispersion forces in layered
systems, the RPA still provides a superior
theoretical framework for making predictions of the interlayer
interactions over the explored length scales, albeit at greater
computational cost.
Since the LDA fails to describe long-range interactions it should be
corrected, whenever possible, to incorporate these effects when
calculating the interlayer elastic and inelastic coefficients, as
demonstrated in this work.

This procedure to assess and improve the qualitative role of the LDA
can be applied routinely to a variety of layered 2D materials. Here we
have used the approach to approximate RPA-level calculations for
bilayers,  that allows to make reliable predictions for interlayer
geometries. 
In order to go beyond the RPA one should consider short-range
correlations by modelling the exchange-correlation kernel that can
incorporate the many-body effects in a more precise
manner~\cite{Olsen2012,doi:10.1063/1.4755286,Dobson2014-1,Jung2004}.

\section{Acknowledgments.}

This work has been supported by the Korean NRF under Grant No. NRF-2016R1A2B4010105 and the Korea Research Fellowship Program through the NRF funded by the Ministry of Science and ICT (NRF-2016H1D3A1023826).
TG acknowledges support of the Griffith University Gowonda HPC Cluster.

\appendix

\section{Parametrization of the potential energy}
\label{app:param}

In the main text, we provide the simple expressions that allow to extract the potential landscape from the energies at AA, AB and BA stacking. Here, we give the more general expressions that allow to extract the information from the combination of \textit{any} three stacking configurations. We omit the $c$ dependence to simplify the notation. After some algebra on Eq.~(\ref{eq:f1}), using the energies of three arbitrary configurations that are given by their respective coordinates
\begin{align}
A = \phi(A_x,A_y) \\ B = \phi(B_x,B_y) \\ C = \phi(C_x,C_y),  
\end{align}
one finds that
\begin{align}
  \phi =& \arctan\left[ \frac{1}{\frac{\delta}{\beta}D-1} \frac{\delta \alpha - \beta \gamma}{\beta \delta} - \frac{\gamma}{\delta}\right],
  \label{eq:phibis}
  \\
  C_1 =& \frac{B-C}{2(\gamma \cos \phi + \delta \sin \phi)}
  \label{eq:C1bis}
\end{align}
and
\begin{multline}
  C_0 = A - 2 C_1 \cos(\phi - G_1 A_y) \\ 
  - 4 C_1 \cos(G_1 A_y/2 + \phi) \cos(\sqrt{3} G_1 A_x/2)
    \label{eq:C0bis}
\end{multline}
where
\begin{equation}
\alpha = a_1 + a_3
\end{equation}
\begin{equation}
\beta = a_2 - a_4
\end{equation}
\begin{equation}
\gamma = b_1 + b_3
\end{equation}
\begin{equation}
\delta = b_2 - b_4
\end{equation}
\begin{equation}
D = \frac{A-B}{B-C}
\label{bigD}
\end{equation}
with
\begin{equation}
a_1 = \cos(G_1 A_y) - \cos(G_1 B_y)
\end{equation}
\begin{equation}
a_2 = \sin(G_1 A_y) - \sin(G_1 B_y)
\end{equation}
\begin{equation}
a_3 = 2 \cos(G_1 A_y/2) m_A - 2 \cos(G_1 B_y/2) m_B
\end{equation}
\begin{equation}
a_4 = 2 \sin(G_1 A_y/2) m_A - 2 \sin(G_1 B_y/2) m_B
\end{equation}
\begin{equation}
b_1 = \cos(G_1 B_y) - \cos(G_1 C_y)
\end{equation}
\begin{equation}
b_2 = \sin(G_1 B_y) - \sin(G_1 C_y)
\end{equation}
\begin{equation}
b_3 = 2 \cos(G_1 B_y/2) m_B - 2 \cos(G_1 C_y/2) m_C
\end{equation}
\begin{equation}
b_4 = 2 \sin(G_1 B_y/2) m_B - 2 \sin(G_1 C_y/2) m_C
\end{equation}
and, finally
\begin{equation}
m_A = \cos(\sqrt{3}G_1 A_x/2)
\end{equation}
\begin{equation}
m_B =  \cos(\sqrt{3}G_1 B_x/2)
\end{equation}
\begin{equation}
m_C =  \cos(\sqrt{3}G_1 C_x/2).
\label{endEqs}
\end{equation}
These expressions have been cross-checked with the simpler expressions
derived previously~\cite{Jung_2015}, and are valid for any system that
possesses trigonal symmetry, as is the case of many layered materials
not considered here.

\section{Bilayer fitting expressions}
\label{app:bi}

The vdW dispersion and the LDA fits for bilayer systems are similar to the expressions in Eqs.~\eqref{eqn:vdW}/\eqref{eqn:vdW2} and \eqref{eqn:ULDA}.
\begin{align}
  U_{\vdW,\text{Bi}}(c)=&-\frac{C_4^\text{Bi}}{(c^2-D_s^{\text{Bi}^2)^2}} - \frac{C_3^\text{Bi}}{c^3} \frac{2}{\pi} \arctan\left(\frac{c}{D_C^\text{Bi}} + \phi_c^\text{Bi}\right)
  \label{eqn:vdWBi}
  \\
  U^{\LDA,\text{Bi}}_{S}(c)=&-M_0^{S,\text{Bi}}\left[\frac{\tau^{S,\text{Bi}}_2 e^{-\tau^{S,\text{Bi}}_1 x_{S}^\text{LDA}} - \tau^{S,\text{Bi}}_1 e^{-\tau^{S,\text{Bi}}_2 x_{S}^\text{LDA}}}{\tau^{S,\text{Bi}}_2 - \tau^{S,\text{Bi}}_1}\right].
  \label{eqn:ULDABi}
\end{align}
with a different set of parameters than the ones obtained for the
bulk. Similar to the bulk, the $C_3^\text{Bi}$-term is non-zero for graphene only. Some of us have argued previously~\cite{Gould_2013} that the
most important changes occur to the parameters $C_3^\text{Bi}$,
$C_4^\text{Bi}$ and $M_0^{S,\text{Bi}}$ and those are the only ones
that have to be rescaled. The method to obtain these scaling factors
for bilayer graphene is outlined in Ref.~[\onlinecite{Gould_2013}],
and are respectively given by 0.455, 0.462 and 0.5. The latter was
obtained assuming $U^{\LDA,\text{Bi}} = \frac{1}{2} U^{\LDA}$. 

Here, we perform the bilayer LDA calculations for all systems, and
confirm that the assumption on $M_0$ is reasonable as a first
approximation. However, some of the G/G bulk structures can,
comparatively, become even more stable in their bilayer form, while the opposite behavior is generally true for the other systems.  
Due to the fact that a bilayer has only one interface there are changes in the interlayer equilibrium distances with respect to bulk. 
In Table~\ref{dataTable}, the results for bilayer have thus been
obtained by directly fitting bilayer LDA data using
Eq.~(\ref{eqn:ULDABi}).

\bibliography{RPA}

\end{document}